\documentclass[acmlarge]{acmart}

\AtBeginDocument{%
  }

\setcopyright{cc}
\setcctype{by}
\acmJournal{IMWUT}
\acmYear{2026} \acmVolume{10} \acmNumber{4} \acmArticle{253}
\acmMonth{12} \acmDOI{10.1145/3857961}

\begin{document}

\title{ElectroCalm: Investigating User Experience and Physiological Responses to Multi-Node Transcutaneous Electrical Stimulation for Stress Modulation}

\author{Yoon Kyung Shon}
\orcid{0009-0004-9133-597X}
\affiliation{
  \institution{University of Maryland} 
  \city{College Park}
  \state{Maryland}
  \country{USA}
}
\email{ykshon@umd.edu}

\author{Pablo E. Paredes}
\orcid{0000-0003-2431-9190}
\affiliation{
  \institution{University of Maryland} 
  \city{College Park}
  \state{Maryland}
  \country{USA}
}
\email{paredes@umd.edu}

\begin{abstract}
Transcutaneous Electrical Stimulation (TES) for stress modulation has largely focused on vagus nerve stimulation at the neck, yet two questions remain underexplored: how users experience electrotactile sensations at this sensitive site, and whether electrotactile feedback itself can function as a haptic stress intervention. 
To investigate this gap, we applied an 8-node TES array to two body sites during a cognitively demanding 2-back task: the neck, a site used in prior vagal TES research for its anatomical proximity to the cervical vagus nerve, and the arm, the most common haptic wearable site, which primarily provides cutaneous electrotactile sensation without vagal access.
In a usability study, both sites were rated as comparably comfortable without disrupting task performance, though the same stimulation was interpreted differently across sites.
A second study assessing the physiological effects of active versus inactive stimulation under identical electrode attachment revealed that only neck TES produced cortical and autonomic changes consistent with stress modulation, while arm TES did not differ from inactive conditions.
These findings suggest that neck TES can be acceptable despite the site's sensitivity, and that both the physiological effects and the sensory experience of TES depend on placement.
\end{abstract}

\ccsdesc[500]{Human-centered computing~Haptic devices}
\ccsdesc[300]{Human-centered computing~Ubiquitous and mobile computing systems and tools}
\ccsdesc[300]{Human-centered computing~Empirical studies in ubiquitous and mobile computing}
\ccsdesc[300]{Applied computing~Consumer health}

\keywords{Stress, Mental Health, Intervention, Haptics, Electrical Stimulation, Electrotactile, User Experience, Wearable}

\maketitle

\section{INTRODUCTION}

Stress is an essential physiological process that supports human survival at appropriate levels~\cite{yu_biofeedback_2018}. 
However, repeated activation of stress responses chronically engages the sympathetic nervous system~\cite{marin_chronic_2011}. Over time, this contributes to physical and mental health problems, including cardiovascular disease, psychological disorders, and impaired work performance~\cite{yaribeygi_impact_2017}. 
This motivates approaches that modulate daily stress responses before they accumulate into chronic strain~\cite{bequet_subtle_2022}, particularly in environments that constitute a significant portion of daily life, such as cognitively demanding workplaces~\cite{howe_design_2022, stansfeld_psychosocial_2006}.

Among such approaches, haptic feedback is well-suited for non-disruptive, everyday use. Previous studies have shown it could operate unobtrusively during cognitive tasks~\cite{paredes_evaluating_2017, brelet_stress_2022, yamane_simple_2021}, as it is often processed with minimal attentional resources~\cite{ro_unconscious_2021}. Studies also demonstrated that haptic feedback can convey comfort and engage parasympathetic responses that reduce physiological stress markers~\cite{eckstein_calming_2020, haynes_calming_2022}.

Within the broader family of haptics, Transcutaneous Electrical Stimulation (TES) offers distinct advantages for stress modulation. 
Unlike mechanical or thermal actuators, TES requires only thin, silent electrodes without moving parts~\cite{patel_latest_2025, vance_using_2022}, and can access autonomic pathways beyond the somatosensory system depending on electrode placement~\cite{howland_vagus_2014}. Studied primarily in neuroscience contexts~\cite{ridgewell_effects_2021, ridgewell_investigation_2022}, TES delivers low-voltage currents through skin electrodes at the neck to target the vagus nerve, a major cranial nerve that regulates autonomic body functions~\cite{verlinden_morphology_2016}.
The neck is therefore the most common TES placement for stress modulation, yet it is also a sensitive site for tactile stimulation~\cite{fleck_wearable_2025}. Combined with the fact that electrotactile stimulation is often perceived as unpleasant in prior HCI research~\cite{delitto_study_1992, faltaous_understanding_2024}, it remains unclear whether users can tolerate and accept electrical stimulation at this sensitive site.

Beyond vagal targeting, TES inherently produces electrotactile sensations through somatosensory pathways~\cite{duente_colorful_2023, emami_survey_2025}.
However, existing TES research has focused almost exclusively on clinical efficacy and vagal mechanisms rather than the experiential and physiological role of the tactile component itself. This gap is notable because other haptic modalities, such as vibrotactile feedback, have demonstrated stress modulation through somatosensory pathways~\cite{kelling_good_2016, bequet_subtle_2022, paredes_calmmenow_2011}. Whether electrotactile feedback can similarly contribute to stress modulation remains an open question for practical deployment where sensory experience shapes adoption~\cite{lee_hapticus_2025, sathiyamurthy_measuring_2021, van_den_berg_facilitators_2025}.

We therefore study two research questions. 
\textbf{RQ1}, how do users perceive and experience TES on the neck, a sensitive site rarely explored in haptic HCI?
\textbf{RQ2}, can electrotactile stimulation at a site without vagal access modulate physiological stress?
%
To that end, we utilized an 8-node electrode array with cross-fading activation patterns inspired by calming sequential haptic patterns~\cite{paredes_evaluating_2017, paredes_just_2018}. By distributing stimulation across multiple nodes, we aimed to reduce single-point electrotactile discomfort while creating smooth spatiotemporal patterns.
We applied this system to two body sites: the neck and the arm. 
The \textit{neck} served as the physiologically targeted TES site, following prior work on autonomic modulation~\cite{howland_vagus_2014}, while its sensitivity to electrotactile stimulation motivates RQ1~\cite{fleck_wearable_2025}.
The \textit{arm} is a conventional haptic wearable site where user experience is well established~\cite{lee_affective_2026}. 
Because arm stimulation is expected to act primarily through cutaneous electrotactile sensation rather than cervical vagal targeting~\cite{ackerley_c-tactile_2022}, this site comparison addresses RQ2.

Using a within-subject design, we assessed user experience, physiological responses including electroencephalography (EEG) and heart rate variability (HRV), and performance on the cognitively demanding 2-back task~\cite{balters_calm_2020, lee_hapticus_2025, duente_colorful_2023}. 
We first conducted a usability study with exploratory physiological assessment at both sites. Both neck and arm TES were rated as comparably comfortable, despite distinct tactile experiences. Task performance remained stable, suggesting that TES did not interfere with cognitive performance. 
Exploratory physiological measures showed promising changes at both sites relative to a device-free control, but could not determine whether effects reflected active stimulation or electrode presence alone. 
We therefore conducted a second study comparing active and inactive stimulation under identical electrode attachment. The two sites dissociated: neck TES produced cortical and autonomic changes consistent with stress modulation, whereas arm TES did not differ from inactive conditions. 
Together, these results suggest that physiological stress modulation appeared specific to neck placement, and that TES can be acceptable on the sensitive neck despite its sensitivity.

\paragraph{\textbf{Contribution}}
This paper makes the following contributions:

\begin{itemize}
\item 
We showed that multi-node spatiotemporal TES can produce acceptable and comfortable experiences on the neck, a sensitive site rarely explored in haptic HCI, rated comparably to the arm.

\item  We demonstrated that neck TES produced consistent cortical and autonomic stress modulation, whereas arm TES did not differ from inactive stimulation, indicating that physiological stress modulation was specific to neck placement.

\item 
We provide design implications for TES stress modulation, including separating physiological targeting from tactile comfort, designing for acceptability at sensitive body sites, and considering how body site shapes the sensory interpretation of stimulation.

\end{itemize}

\section{RELATED WORK}

\subsection{TES for Stress Modulation}

Stress responses activate the sympathetic nervous system, leading to physiological changes including elevated heart rate, increased cortisol levels, and altered autonomic activity~\cite{marin_chronic_2011, yaribeygi_impact_2017}. 
To address these autonomic responses, researchers have increasingly investigated direct nerve stimulation as a targeted approach to attenuate stress states~\cite{prasad_impact_2025}, with TES emerging as a promising non-invasive technique that delivers controlled electrical pulses through skin electrodes to stimulate underlying neural pathways~\cite{vance_using_2022, patel_latest_2025, ridgewell_investigation_2022}. 
TES uses low-voltage currents that can be precisely modulated in frequency, intensity, and pulse width to target specific neural structures~\cite{hsieh_ultra-low_2021}, and is low-risk, cost-effective, and adaptable to various body sites for therapeutic applications~\cite{vance_using_2022}.

Originally developed for pain relief by interrupting pain signals traveling to the brain~\cite{rushton_electrical_2002}, recent TES research has expanded to stress modulation by applying stimulation to head and neck regions~\cite{fertonani_transcranial_2017}. More commonly, researchers have focused on vagus nerve stimulation, as this major cranial nerve is the primary driver of the parasympathetic nervous system~\cite{howland_vagus_2014}, playing a critical role in downregulating sympathetic activity~\cite{ridgewell_effects_2021}. 
Extending from the brainstem through the neck to organs in the chest and abdomen~\cite{verlinden_morphology_2016}, the vagus nerve serves as a key pathway for restoring autonomic balance. 
These applications typically use single-node electrode systems placed on or around the neck or ear area~\cite{ridgewell_effects_2021, ridgewell_investigation_2022} and have demonstrated potential to reduce physiological stress responses and enhance cognitive performance~\cite{van_dijk_effects_2002}.

However, this line of work has primarily evaluated TES as an intervention, with efficacy measured through physiological outcomes. The electrotactile sensation that TES inherently produces at the skin surface, and how users experience, tolerate, and accept it, has received far less attention. This gap matters because stress modulation wearables depend not only on physiological efficacy but also on whether users find the stimulation comfortable and acceptable enough to wear repeatedly~\cite{van_den_berg_facilitators_2025, lee_hapticus_2025, sathiyamurthy_measuring_2021}. Without explicit attention to user experience, prior TES findings offer limited guidance for wearable TES design.

\subsection{TES as Electrotactile Feedback}

In parallel with neuromodulation applications, TES has been used in HCI as electrotactile feedback, providing sensory experiences via somatosensory pathways~\cite{hsieh_ultra-low_2021, johnson_transcutaneous_2007, shon_when_2026}. On activation, electrical stimulation predominantly recruits cutaneous A$\beta$ afferents to encode touch sensations such as vibration or pressure, rather than C-tactile (CT) afferents, which are unlikely to be directly activated by electrical stimulation delivered through the skin at comfortable levels~\cite{ackerley_c-tactile_2022}. 
This makes electrotactile well suited for producing diverse tactile experiences, while also implying that any affective or calming effects of electrotactile feedback must arise through A$\beta$-mediated or higher-level cognitive routes.

These electrotactile capabilities have enabled diverse applications, from virtual reality and haptic displays that augment surface contact perceptions~\cite{zhou_electrotactile_2022, kourtesis_electrotactile_2022} to non-disruptive guidance and communication systems~\cite{alotaibi_use_2020}.
However, electrotactile sensations present significant user experience challenges, as perceived sensations vary across users and sites, and are often reported as ranging from pleasant tapping to uncomfortable prickling or shocking~\cite{duente_colorful_2023, van_schaik_verbal_2026}.
In particular, single-node stimulation systems are often associated with discomfort that limits extended use~\cite{delitto_study_1992, faltaous_understanding_2024}.
Prior work has begun to address these comfort limitations along two complementary directions. The first is parameter tuning, examining how stimulation parameters such as intensity, frequency, and pulse width shape perceived acceptability~\cite{alotaibi_first_2022, duente_colorful_2023}. 
The second is the spatial and temporal distribution of stimulation, where interleaved activation across multiple surface electrodes has been shown to reduce localized sensations near individual electrodes and produce more comfortable percepts compared to single-channel stimulation~\cite{pena_channel-hopping_2021}. Nevertheless, substantial individual variability in how users experience electrotactile stimulation remains, particularly across different body sites, and electrotactile sensation itself has rarely been investigated as a stress modulation modality. Little is known about whether electrotactile feedback can be perceived as comfortable during stress, whether it can influence stress states, or how different body sites shape this experience.

\subsection{Haptic Interventions for Stress Modulation}

\begin{table}[t]
\centering
\footnotesize
\setlength{\belowcaptionskip}{6pt}
\renewcommand{\arraystretch}{1.5}
\begin{tabular}{p{2.7cm} p{2.7cm} p{2.7cm} p{2.7cm} p{3.2cm}}
\toprule
\textbf{Feature} & \textbf{Vibrotactile} & \textbf{Thermal} & \textbf{Pressure-based} & \textbf{TES} \\
\midrule

Primary pathway &
Somatosensory \newline (mechanoreceptors) &
Somatosensory \newline (thermoreceptors) &
Somatosensory \newline (deep pressure) &
Somatosensory + \newline Autonomic (site-dependent) \\

Form factor &
Bulky (motor-based) &
Rigid (Peltier) &
Bulky (pneumatic) / \newline Thin-wire (SMA) &
Thin, flexible \newline (electrodes) \\

Actuation mechanism &
Mechanical oscillation &
Thermal change &
Inflation (pneumatic) / \newline Contraction (SMA) &
Electrical stimulation \\

Spatial control &
Multi-node capable \newline (diffuse perception) &
Multi-node capable \newline (thermal diffusion) &
Area-based control \newline (distributed force) &
Multi-node array \newline (higher spatial selectivity) \\

Auditory footprint &
Audible &
Silent &
Audible (pneumatic) / \newline Silent (SMA) &
Silent \\

\bottomrule
\end{tabular}
\caption{High-level comparison of haptic modalities for stress modulation. TES is distinguished by its ability to access autonomic pathways depending on electrode placement, together with a thin, silent form factor. Pressure-based approaches include pneumatic and shape-memory alloy (SMA) actuators.}
\Description{Comparison of haptic modalities for stress modulation.}
\label{tab:design-space}
\end{table}

As the sense of touch can engage parasympathetic responses that reduce stress markers such as heart rate, blood pressure, and cortisol~\cite{eckstein_calming_2020, haynes_calming_2022}, haptic stress interventions have employed diverse modalities (Table~\ref{tab:design-space}), each offering different trade-offs between physiological efficacy, tactile comfort, and practical deployment.
Vibrotactile feedback represents the most extensively studied approach, with implementations ranging from simple pulsing patterns to complex spatiotemporal sequences~\cite{kelling_good_2016, bequet_subtle_2022, paredes_calmmenow_2011}. 
These systems primarily activate skin mechanoreceptors to create sensations of vibration or movement, with demonstrated stress-relief effectiveness across various contexts~\cite{paredes_just_2018, balters_calm_2020, yamane_simple_2021}. 
Yet vibrotactile approaches face practical limitations, as sensations can be masked by environmental vibrations and mechanical components often generate audible noise~\cite{alotaibi_first_2022, ryu_vibrotactile_2010}.
Also, vibration diffuses across the skin~\cite{cholewiak_vibrotactile_2003}, limiting the density of spatiotemporal patterns in compact body areas.

Other haptic stress interventions use pressure-based or thermal feedback. 
Pneumatic systems use inflatable cuffs or air chambers to provide dynamic pressure patterns~\cite{vaucelle_design_2009, haynes_calming_2022, choi_aspire_2021}, fostering comfort and a sense of security~\cite{brelet_stress_2022, zhao_affective_2022}, but typically require bulky components. 
Shape-memory alloy actuators offer thinner form factors through contractile wire mechanisms~\cite{compton_dynamic_2021, foo_soft_2020}, but may introduce thermal safety concerns. 
Thermal feedback modulates skin temperature through Peltier elements~\cite{he_affective_2024, umair_exploring_2021}, but demands substantial power and continuous calibration due to sensitivity to ambient temperature~\cite{ju_thermal_2022}. 
Unlike these modalities, which primarily operate through somatosensory pathways, TES provides tactile sensations while also accessing autonomic pathways depending on electrode placement~\cite{patel_latest_2025, vance_using_2022}. 
Furthermore, thin, low-power electrode arrays enable compact, silent operation without moving parts and can be embedded into textiles for everyday wear~\cite{pamungkas_overview_2018}, positioning TES as a distinctive modality in the haptic stress modulation design space.

Despite this diversity, existing work has concentrated on a small set of body sites. A recent systematic review of affective wearable haptic studies found that the forearm (31\%), wrist (18\%), and hand (17\%) together accounted for roughly two-thirds of all placements, while sites such as the neck appeared only rarely~\cite{lee_affective_2026}. 
This distribution reflects established wearable design conventions, where familiar placements such as the forearm are preferred~\cite{van_den_berg_facilitators_2025}, leaving limited empirical evidence on whether physiologically relevant but sensitive placements can be acceptable for stress modulation systems~\cite{fleck_wearable_2025}. 
We therefore investigate whether electrotactile stimulation on the neck can be made comfortable and acceptable during a cognitively demanding stress task.

\section{TES SYSTEM DESIGN}

\begin{figure}[t]
    \centering
    \includegraphics[width=\linewidth]{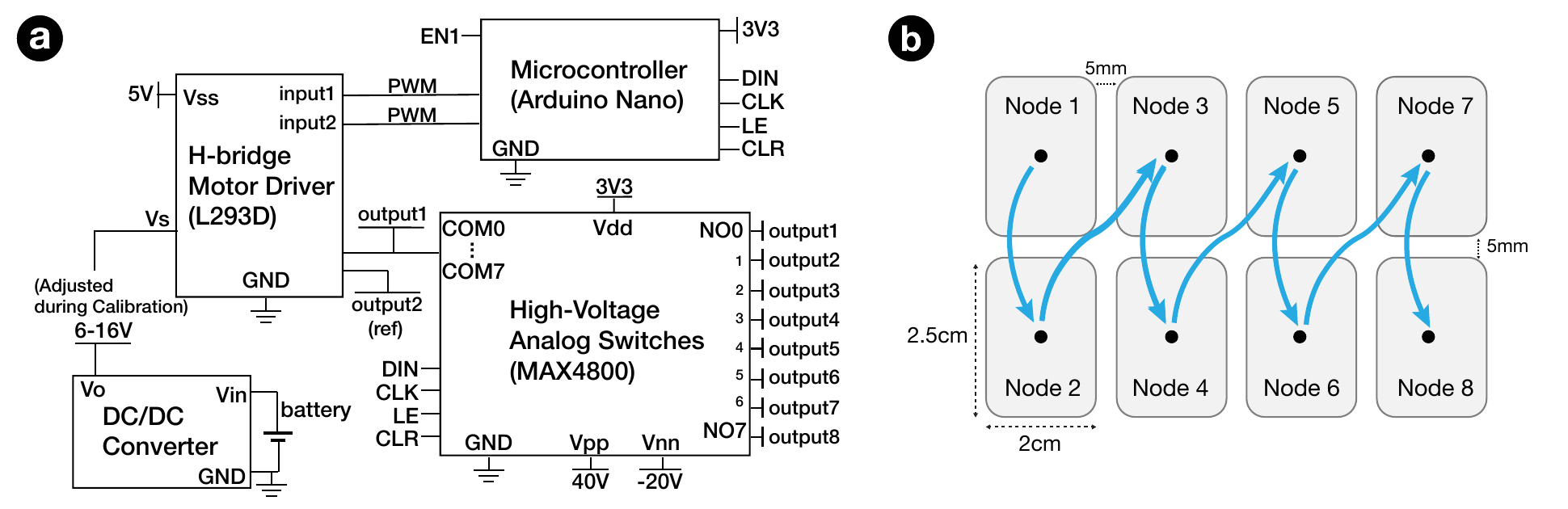}
    \caption{TES System design and stimulation pattern. 
    (a) Hardware schematic of the TES system. 
    (b) Electrode array arranged in a $2\times4$ configuration ($2\times2.5$\,cm pads, 5\,mm spacing), with activation progressing sequentially from node 1 to node 8.}
    \Description{System design and stimulation pattern}
    \label{fig:TES}
\end{figure}

\subsection{Multi-Node TES Device}

We designed a voltage-controlled TES device with an 8-channel output system that enables programmable stimulation patterns to create comfortable electrotactile sensations.
We selected voltage control because it enabled a compact, low-power H-bridge implementation for our wearable, multi-node system. Although current-controlled stimulation is preferred for its robustness to skin-impedance variation in clinical settings~\cite{merrill_electrical_2005} and high-fidelity electrotactile systems~\cite{withana_tacttoo_2018, sakamoto_displaying_2025}, implementing current control in our architecture would require additional current-sensing and feedback circuitry, increasing component count and complexity.
In light of these considerations, we calibrated stimulation intensity separately for each participant and site to account for baseline impedance differences and used short (2-minute) stimulation blocks to reduce the potential impact of within-session impedance drift~\cite{yem_prediction_2025}.

Building on the work of Nishida et al.~\cite{nishida_biosync_2017}, our system incorporates additional components to support smooth, multi-node stimulation across distributed arrays. 
As shown in Figure~\ref{fig:TES}a, the system includes an Arduino Nano microcontroller for pulse generation and timing, a MAX4800 high-voltage analog switch for multiplexing stimulation across eight electrodes, and an L293D H-bridge driver for polarity modulation. 
This configuration supports biphasic pulse delivery to reduce net charge accumulation and is optimized for wearable applications~\cite{emami_survey_2025}. Power is supplied through DC-DC converters, delivering a regulated output up to 35\,V peak.
To ensure safe and comfortable stimulation, we calibrated intensity to a just-perceptible level, which kept the delivered voltage well below the peak capacity (see \S\S~\ref{study1_conditions} and~\ref{study2_conditions}).

\subsection{Spatiotemporal TES Pattern}

To address the discomfort commonly associated with single-node electrical stimulation~\cite{delitto_study_1992, faltaous_understanding_2024}, we designed a multi-node spatiotemporal pattern that distributes activation across eight electrodes~\cite{jingu_shaping_2024, suga_softness_2023}. This design was inspired by research showing that directional touch patterns can evoke pleasant tactile experiences~\cite{zhao_affective_2022, israr_towards_2018}.

Our TES system employed $2\times2.5$\,cm electrodes arranged in a $2\times4$ array (Figure~\ref{fig:TES}b). 
We used a spatial configuration with array pads as distributed active electrodes and a single larger reference electrode as the common return.
Each electrode delivered 2\,Hz biphasic pulses of 400\,$\mu$s width, with the biphasic pulse delivery following conventional TES safety parameters~\cite{patel_latest_2025}. 
Sequential node activation created directional touch patterns across the skin at approximately 4\,cm/s, chosen to provide comfortable touch speeds from prior work~\cite{he_affective_2024, loken_coding_2009}.
PWM-based amplitude modulation (20-80\% duty cycle) enabled smooth crossfading between adjacent electrodes to prevent abrupt sensations~\cite{asplund_its_2020}.
This configuration distributes stimulation rather than concentrating it on single points, reducing localized discomfort while creating a moving electrotactile sensation.

In addition, we calibrated stimulation intensity to a just-perceptible level for each participant, intentionally keeping sensation in a subtle range. This design choice targets a recurring challenge in electrotactile research that stimulation at higher intensities is frequently reported as unpleasant~\cite{delitto_study_1992, faltaous_understanding_2024}, which undermines both user comfort and long-term adoption~\cite{murnane_mobile_2015}. 
By combining spatial distribution across multiple nodes with just-perceptible intensity, we aimed to produce electrotactile sensations that users could experience as comfortable or even pleasant rather than as aversive.

\section{Study~1: Usability Study with Exploratory Physiological Assessment}

We first conducted a usability study to evaluate the multi-node TES system at the arm and neck during a cognitively demanding task, with exploratory physiological measures to assess whether TES could produce observable stress modulation. The studies in this paper were approved by the University Institutional Review Board (IRB 2187447).

\subsection{Conditions} \label{study1_conditions}

\begin{figure}[t]
    \centering
    \includegraphics[width=\linewidth]{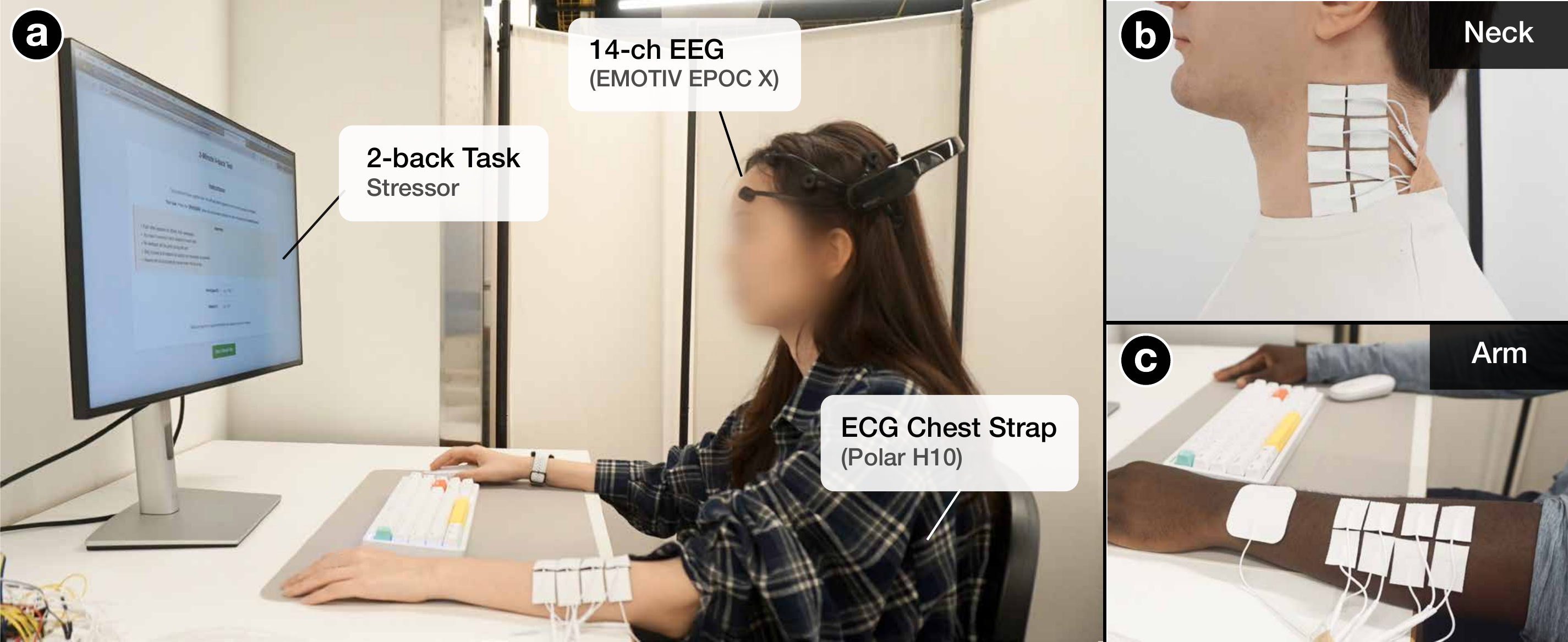}
    \caption{(a) Study setup: participant performs the 2-back task with a 14-channel EEG headset and HRV chest strap worn underneath the shirt. 
    (b) \textit{Neck} condition: TES electrode array placed on the left lateral neck. The reference node is hidden underneath the t-shirt. 
    (c) \textit{Arm} condition: TES electrode array on the non-dominant forearm. The larger electrode serves as a common reference node.}
    \Description{Apparatus showing participant setup, neck electrode placement, and arm electrode placement.}
    \label{fig:apparatus}
\end{figure}

\paragraph{\textbf{Neck}}
Electrodes were placed on the lateral cervical region as presented in Figure~\ref{fig:apparatus}b, providing electrotactile sensation near the cervical vagus nerve, which courses through this area before branching to thoracic and abdominal organs~\cite{verlinden_morphology_2016, howland_vagus_2014}. Consistent with clinical practice, we used the left side to minimize potential cardiac risks, as right-side cervical stimulation more directly affects heart rhythm~\cite{howland_vagus_2014}.

\paragraph{\textbf{Arm}}
Electrodes were placed on the forearm to deliver electrotactile sensation (Figure~\ref{fig:apparatus}c) through somatosensory pathways. 
At comfortable intensities, this placement primarily activates cutaneous A$\beta$ afferents and is not expected to recruit CT afferents or directly target vagal pathways~\cite{ackerley_c-tactile_2022}.
We therefore used the arm as a placement that primarily provides the electrotactile sensation to examine whether such input alone modulates stress.

\paragraph{\textbf{Control}}
To establish a device-free baseline, the control condition excluded electrode attachment during the stressor task. This allowed us to evaluate TES effects on both physiological responses and subjective experience relative to an unencumbered baseline.

For the TES conditions, stimulation at both body sites was calibrated for each participant through interactive communication, gradually increasing voltage from the lowest setting until a just-perceivable sensation was reported and confirmed, establishing this as intensity level 1 on a 1-10 scale~\cite{dolker_perception_2022}. 
This approach normalized subjective intensity across sites, addressing varying skin properties while ensuring low-salience stimulation. The calibrated intensities were \textit{arm} ($M=10.27~\mathrm{V}$, $SD = 2.96$) and \textit{neck} ($M = 8.20~\mathrm{V}$, $SD = 1.19$).

\subsection{Participants}

We recruited 24 eligible participants from the local community (9 female and 15 male, $M=28.29~years$, $SD=10.91$), a sample size comparable to previous HCI physiological studies~\cite{valente_modulating_2024, paredes_just_2018, balters_calm_2020}.
Prior to enrollment, we screened potential participants to exclude contraindications for TES, including cardiac pacemakers, metal implants, epilepsy, pregnancy, skin sensitivities, and current anxiety or other mental health conditions. On the day of the study, participants confirmed they had not consumed caffeine or medication. 
All participants provided written informed consent, including permission for audio recordings and transcription of comments, and received \$20 USD as compensation.

\subsection{Apparatus}

The study took place in a quiet, partitioned room with participants seated facing a monitor (see Figure~\ref{fig:apparatus}a). 
Participants wore an Emotiv EPOC X EEG headset~\cite{emotiv_emotiv_2025} with verified signal quality~\cite{konakoglu_effect_2023, dinh_neuresonance_2025} and a Polar H10 chest strap for HRV~\cite{polar_electro_polar_2025}, with the chest strap worn underneath their clothes.

\subsection{Task \& Procedure}

\begin{figure}[t]
    \centering
    \includegraphics[width=\linewidth]{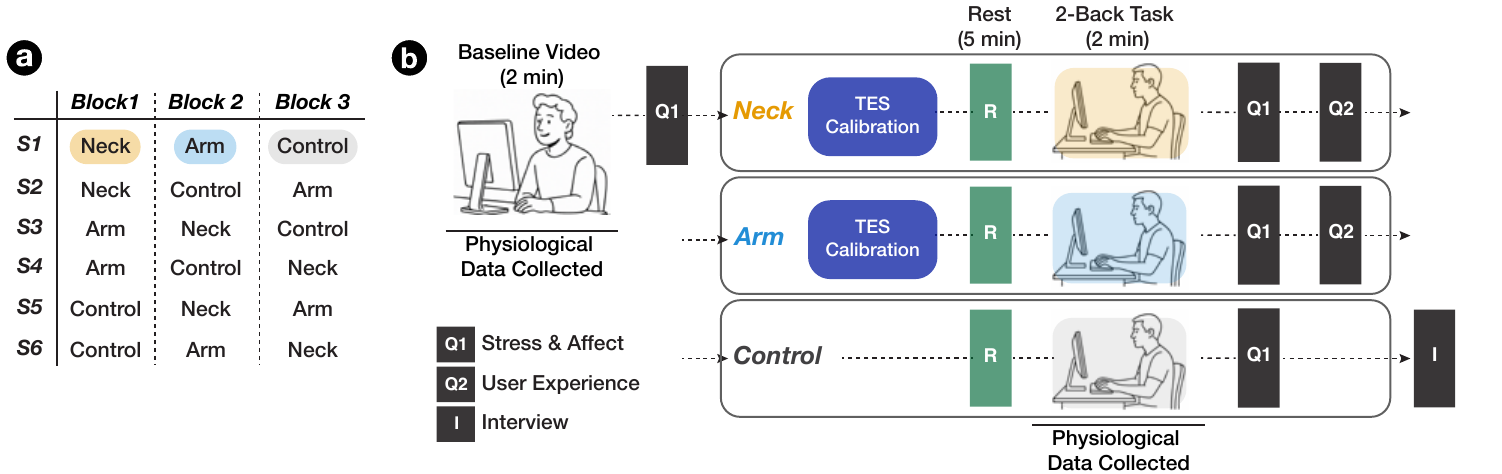}
    \caption{Study~1 procedure: (a) The six counterbalanced sequences (S1--S6) across three blocks. 
    (b) Example sequence S1, where Block 1 is Neck, Block 2 is Arm, and Block 3 is Control, with questionnaire labels below. 
    The session began with a 2-minute baseline video with physiological recording, followed by a stress and affect questionnaire (Q1). Each condition block included TES calibration (TES conditions only), a 5-minute rest period, and a 2-minute 2-back task with continuous physiological recording. Post-task measures included Q1 after all conditions and Q2 for user experience after TES conditions only. The session concluded with a semi-structured interview.}
    \Description{Study~1 procedure: (a) The six counterbalanced sequences (S1--S6) across three blocks. (b) Example sequence S1 showing baseline video, three condition blocks with TES calibration, rest periods, 2-back tasks, questionnaires, and a concluding interview.}
    \label{fig:procedure}
\end{figure}

We employed the \textit{n}-back task, a well-established working memory task that reliably induces acute stress and cognitive load~\cite{meule_reporting_2017}. 
In this task, participants pressed the spacebar when the current stimulus matched the one presented \textit{n} steps earlier. 
We employed a 2-back task as a stressor~\cite{valente_modulating_2024}, which elicits measurable stress responses yet remains manageable for participants throughout the session.

Before beginning the study, participants received an overview of the TES device, stimulation procedures, and electrode placement sites (neck and arm), and were not informed of the specific physiological mechanisms for each placement. Participants were required to remove metallic accessories and were informed they could request immediate cessation of stimulation or withdraw from the study at any time.
After providing consent, participants completed two practice blocks of the 2-back task with only 15 letters each to familiarize themselves with the task. The main study then followed the protocol illustrated in Figure~\ref{fig:procedure}. 
For baseline measurements, participants watched a 2-minute nature video~\cite{grassini_watching_2022, zhao_affective_2022} and completed the perceived stress and affect questionnaire.

The experimental phase consisted of three conditions: \textit{control} (no stimulation), \textit{neck}, and \textit{arm}. 
To control for potential order effects, all three conditions were fully counterbalanced across participants. 
After each condition, participants completed the perceived stress and affect questionnaire. Following the \textit{arm} and \textit{neck} conditions, participants also completed additional stimulation experience ratings. 
A 5-minute rest interval between conditions minimized carryover effects, reduced fatigue, and allowed participants to habituate to electrode sensation, minimizing initial arousal or orienting responses to the device~\cite{prasad_impact_2025, ridgewell_investigation_2022}.
Throughout the study, an investigator continuously monitored participants for any signs of discomfort or adverse effects.

\subsection{Measures}

\subsubsection{\textbf{User Experience and Subjective Response}}

We assessed subjective responses after each condition using self-reports of perceived stress~\cite{mengelkoch_using_2023, murray_psychometric_2023}, valence, and arousal~\cite{balters_calm_2020, brelet_stress_2022, alotaibi_investigating_2020} to evaluate the influence of the stimulation. 
Stress was measured on a 0–10 scale (0 = not stressed at all, 10 = extremely stressed), whereas valence and arousal were measured on a 1–9 scale (1 = unpleasant, 9 = pleasant for valence; 1 = calm, 9 = excited for arousal). All measures were baseline-corrected to isolate the effects of each condition.
Because our system was designed for tactile comfort during stress rather than conscious affect modulation, valence and arousal served as supplementary measures to assess whether participants perceived any affective changes during TES.

Following the TES stimulation, participants also completed ratings related to haptic feedback experiences~\cite{sathiyamurthy_measuring_2021, lee_hapticus_2025}, sensations specific to TES~\cite{ridgewell_investigation_2022, duente_colorful_2023}, and user perceptions of tactile stimulation~\cite{choi_aspire_2021}, after the \textit{arm} and \textit{neck} placement, all measured on a 7-point scale (1 = strongly disagree, 7 = strongly agree). 
Specifically, participants rated the stimulation's pleasantness, tactile comfort, subtlety, focus on task, wearability for extended periods, and willingness for future use. We also included open-ended questions asking participants to describe how the stimulation felt. 
Finally, semi-structured interviews were recorded and transcribed. The author conducted reflexive thematic analysis~\cite{bager-charleson_supporting_2022}, performing initial open coding and iteratively refining themes, which were then mapped to questionnaire dimensions.

\subsubsection{\textbf{Task Performance}}
We evaluated participants' 2-back task performance using accuracy and response time~\cite{meule_reporting_2017, valente_modulating_2024}. 
In this task, correct responses included hits (pressing the spacebar when the current letter matched the one from two steps earlier) and correct rejections (withholding a response when no match occurred), while errors comprised false alarms (responding without a match) and misses (failing to respond to valid matches).
Accuracy was calculated as the proportion of correct responses (hits + correct rejections) to total trials. Response time was calculated as the mean reaction time for correct target responses (hits only).

\subsubsection{\textbf{Exploratory Physiological Assessment}}

We monitored EEG and HRV, commonly used physiological measures for capturing responses to stress~\cite{balters_calm_2020, he_affective_2024, de_looff_associations_2018}, as in prior TES research where direct vagal recording is not feasible non-invasively~\cite{ridgewell_investigation_2022, yildirim_eeg_2020, vos_stress_2025, konakoglu_effect_2023}.
Because of the large inter-individual variability in baseline physiological levels, we analyzed responses relative to each participant's baseline condition measured during the video-watching phase (see Figure~\ref{fig:procedure})~\cite{zhao_affective_2022}.
We also collected electrodermal activity (EDA) via a wrist sensor, but excluded it from the analysis because continuous keyboard input during the 2-back task introduced motion artifacts that confounded the signal.

\paragraph{\textbf{Electroencephalography (EEG)}}
EEG has been widely used for stress monitoring due to its temporal sensitivity~\cite{premchand_wearable_2024, vos_stress_2025}, and can assess cortical responses to tactile stimulation~\cite{ridgewell_investigation_2022}. 
We employed a wireless 14-electrode EEG device, shown in the study setup in Figure~\ref{fig:apparatus}a, with the electrode layout illustrated in Figure~\ref{fig:topo}a, and validated it for reliable brain activity measurement~\cite{pigtain_effectiveness_2024}. 
EEG data were sampled at 128\,Hz, collected in real-time using the Lab Streaming Layer protocol. We focused on alpha ($\alpha$: 8-13\,Hz) and beta ($\beta$: 13-30\,Hz) band power on frontal channels (AF3, AF4, F3, F4)~\cite{yildirim_eeg_2020, vos_stress_2025}, where increased alpha activity is linked to relaxed states and elevated beta power to greater cognitive engagement.

Raw EEG signals were preprocessed following established methods~\cite{bigdely-shamlo_prep_2015, pedroni_automagic_2019}. 
Data epochs were first segmented with the initial and final 10 seconds discarded to eliminate wrap-up segments. 
A band-pass filter of 0.5-50\,Hz was then applied and segmented into 2-second epochs. Artifact rejection was applied using a moving window peak-to-peak voltage threshold (100-300\,$\mu$V).
Spectral power was extracted using Welch's method to compute power spectral density across frequency bands, log-transformed to the decibel scale, and baseline-corrected to each participant's \textit{baseline} state values~\cite{lan_real-time_2016}.

\paragraph{\textbf{Heart Rate Variability (HRV)}}

Heart Rate Variability (HRV) is a non-invasive marker of autonomic nervous system activity~\cite{rajendra_acharya_heart_2006} that reflects the variation in time intervals between consecutive heartbeats. 
Because stress reduction interventions typically enhance parasympathetic activity, which counteracts sympathetic arousal and promotes physiological calm~\cite{de_looff_associations_2018}, HRV is widely used to index stress-related autonomic regulation~\cite{pham_heart_2021, kim_stress_2018}.

We monitored heart activity using a Polar H10 chest-strap electrocardiogram (ECG) sensor~\cite{polar_electro_polar_2025} at a sampling rate of 130Hz, with R–R intervals extracted via Elite HRV~\cite{moya-ramon_validity_2022, elite_hrv_best_2025}. 
Given the 2-minute measurement windows, we focused on the Root Mean Square of Successive Differences (RMSSD) and Heart Rate (HR), both recommended for short-duration HRV analysis~\cite{shaffer_overview_2017, salahuddin_ultra_2007}. RMSSD indexes parasympathetic activity, with higher values indicating reduced stress, while HR captures overall autonomic balance, with higher values indicating increased stress.
For each participant, values were baseline-corrected and z-scored to minimize between-subject variability.

\subsection{Findings}

All participants completed all three conditions without adverse events or early termination. We report findings in three parts: user experience as the primary outcome, followed by task performance and subjective response, and finally, exploratory physiological assessments across conditions.

\subsubsection{User Experience}
\begin{figure}[t] 
    \centering
     \includegraphics[width=\linewidth]{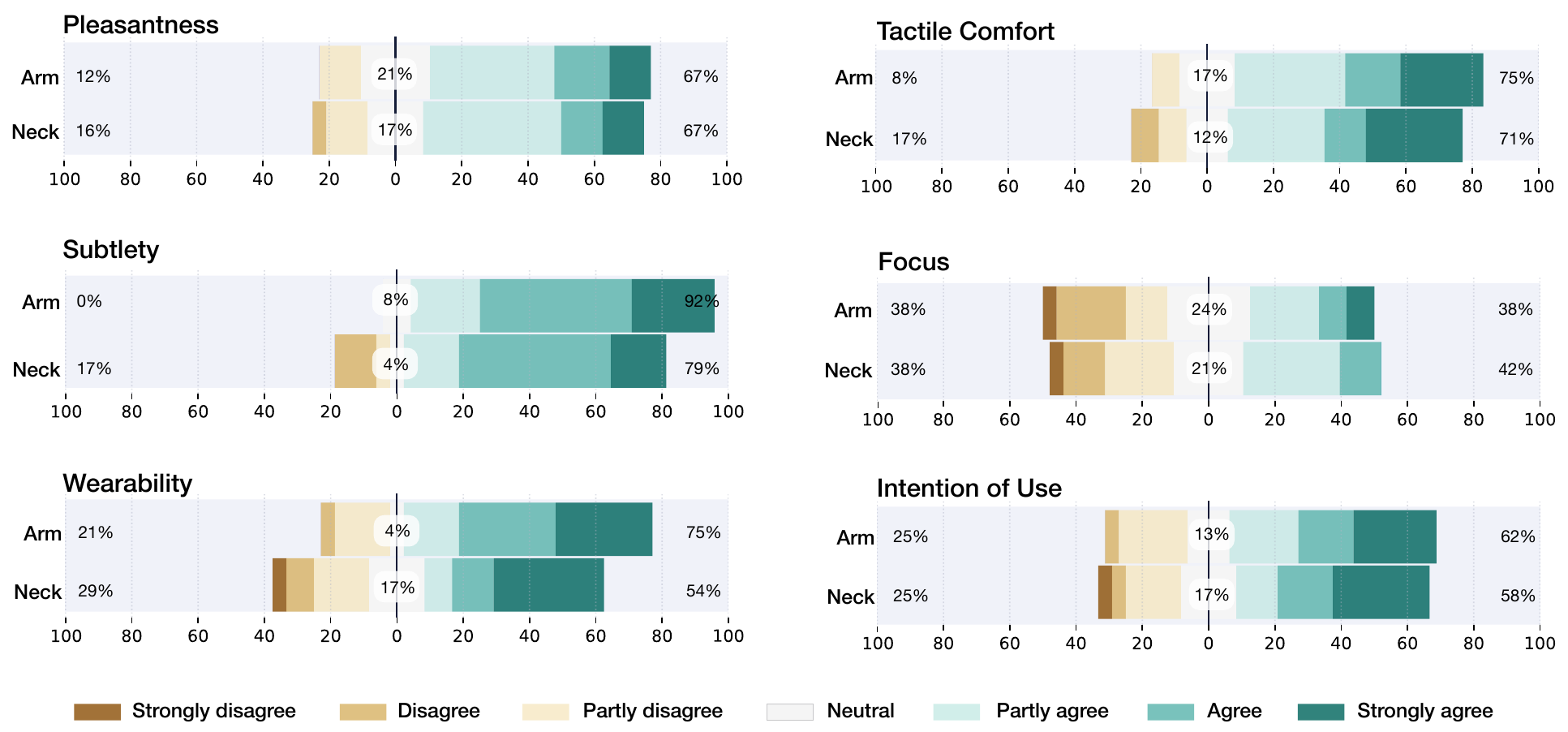}
     \caption{Diverging 100\% stacked-Likert summaries comparing \textit{arm} and \textit{neck} TES placement across six metrics. Each row shows the percentage distribution of 7-point responses, disagreement to the left, neutral centered, and agreement to the right. In-line numbers indicate the shares of negative, neutral, and positive responses for each row.}
    \Description{Stacked-Likert summaries comparing arm and neck TES placement across six metrics.}
    \label{fig:q2_result}
\end{figure}

Participants reported distinct experiences between \textit{arm} and \textit{neck} placement across multiple perceptual dimensions. 
These reports are illustrated in Figure~\ref{fig:q2_result}, and we focus on descriptive patterns and qualitative interview data that characterize the two sites.

\paragraph{\textbf{Stimulation Experience}}
Participants described distinct sensory experiences between body sites. \textit{Arm} placement was predominantly characterized as familiar tactile sensations such as tapping or pulsing (11/24), buzzing (4/24), or gentle vibration (2/24), described as \textit{"like tapping, not uncomfortable"} (P6) or \textit{"ants walking on the skin"} (P16). 
\textit{Neck} placement evoked more varied and metaphorical responses beyond typical tactile descriptions, with pulsating sensations described as \textit{"very gentle throbbing"} (P14) and contact metaphors such as \textit{"raindrops"} (P7), \textit{"pebble hitting a lake"} (P8) or \textit{"butterflies in your stomach, but instead on your neck"} (P2).

Notably, participants frequently reported sensations of social presence, particularly with \textit{arm} placement (5/24), describing experiences such as \textit{"someone having a hand on your arm"} (P24), \textit{"like someone saying it's ok, it's ok"} (P7), or \textit{"someone holding my arm and saying relax, calm down"} (P17). 
The \textit{neck} elicited fewer but similar responses (2/24), such as \textit{"someone's next to me"} (P19).

\paragraph{\textbf{Tactile Pleasantness and Comfort}}

Both body sites received moderate \textbf{pleasantness} ratings, with comparable scores for the \textit{arm} $(M=4.96, SD=1.20)$ and \textit{neck} $(M=4.83, SD=1.31)$. Participants (9/24) often described \textit{arm} placement as soothing, with comments like \textit{"felt like soothing ASMR"} (P7) or \textit{"a patting"} (P17).
For the \textit{neck}, some participants (6/24) found it pleasant or calming, while others (4/24) described it as uncomfortable, e.g., \textit{"It made me more nervous"} (P23). 
Many (10/24) characterized both sites as mildly pleasant or neutral, with a few linking the sensations to symbolic experiences, e.g., \textit{"it reminds me of childhood plays that induced goosebumps"} (P8).

\textbf{Comfort} perceptions were similar across locations with \textit{arm} ($M=5.33, SD=1.27$) and \textit{neck} ($M=5.17, SD=1.61$). Many participants (13/24) indicated that \textit{arm} placement felt more natural and less intrusive, often linking it to familiar experiences such as wearing watches, e.g., \textit{"because I've received stimulation in the hands or arm before, but never in the neck"} (P12). By contrast, the \textit{neck} was more often associated with discomfort or intrusiveness (6/24), being described as \textit{"felt more intrusive, because of tickles"} (P12), \textit{"it's a more vulnerable location"} (P14), or even \textit{"just felt un-normal"} (P19). However, many participants (11/24) still perceived \textit{neck} placement as reasonably natural, though some expressed concerns about sensitivity, e.g., \textit{"Is it touching my blood vessel?"} (P16).

\paragraph{\textbf{Subtlety and Focus}}

Participants rated the \textit{arm} placement as slightly more \textbf{subtle} $(M=5.88, SD=0.90)$ compared to the \textit{neck} $(M=5.29, SD=1.57)$. 
A considerable number of participants (8/24) described the stimulation on both sites as subtle, ambient, or ignorable, e.g., \textit{"ambient background noise"} (P3) or \textit{"helped me cut through the fog like background music"} (P22). However, opinions varied on which site was more noticeable, with some indicating the \textit{neck} was easier to ignore (10/24), e.g., \textit{"the arm was more active"} (P24), while others found the \textit{arm} more subtle (6/24), e.g., \textit{"the neck I was actively aware of"} (P12).

Ratings for perceived \textbf{focus} clustered around the neutral point, with both sites receiving identical mean ratings for the \textit{arm} $(M=3.96, SD=1.65)$ and \textit{neck} $(M=3.96, SD=1.40)$. 
Most participants (14/24) reported that stimulation at both placements allowed them to remain fully focused on the task, while others (10/24) noted divided attention, where several (5/24) described the \textit{arm} as more distracting, e.g., \textit{"distracting and unpredictable"} (P5), while a smaller number (3/24) pointed to the \textit{neck}, e.g., \textit{"I was actively aware of the stimulation on the neck"} (P9). A few (3/24) found that both sites affected their focus, e.g., \textit{"I could focus better when there was no stim"} (P23). Interestingly, some participants (4/24) described the stimulation as supportive for their task, noting it was \textit{"alerting but not distracting"} (P3) or served as \textit{"a reminder to focus"} (P22).

\paragraph{\textbf{Wearability, Future Use, and Site Preference}}

\textit{Arm} placement was more favorable in terms of \textbf{wearability}, with participants indicating they could wear the device longer without discomfort in the \textit{arm} ($M=5.38, SD=1.58$) than \textit{neck} ($M=4.88, SD=1.96$). 
Many participants (13/24) found the \textit{arm} more natural and familiar, likening it to wearing a watch or band, and noted it could be easily hidden under clothing and felt generally less intrusive, e.g., \textit{"I'm used to wearing things on my arm"} (P1) or \textit{"The neck is visible, the arm can be covered"} (P16). 
The \textit{neck} was sometimes preferred (4/24) for not interfering with arm and hand use during work-like activities, e.g., \textit{"On the neck it won't interfere with work"} (P10). However, sensitivity and intrusiveness concerns were more common for the \textit{neck} (5/24), with participants describing it as tickly, vulnerable, or odd, and some mentioned adhesive awareness, e.g., \textit{"it would be weird if I have something on my neck"} (P6).

When asked about intention for \textbf{future use}, responses were nearly identical in both sites, with the \textit{arm} $(M=5.00, SD=1.62)$ and the \textit{neck} $(M=4.96, SD=1.83)$. 
While most participants (17/24) expressed willingness to use the device on either site, others showed conditional acceptance, indicating they would use the stimulation if it were more discreet or showed clear benefits, e.g., \textit{"As long as it doesn't interfere with daily life"} (P16) or \textit{"if it helps me, I'll wear it"} (P18). Some (4/24) also mentioned they would choose based on effectiveness, selecting the site where they felt clearer or more focused, e.g., \textit{"I think I did better with neck stimulation"} (P6).

Overall, participants showed divided \textbf{preferences}, with \textit{arm} slightly favored (13/24) over \textit{neck} (11/24).
\textit{Arm} preference was attributed to familiarity, comfort, unobtrusiveness, and lower perceived intrusiveness, with comments like "more natural feeling to have" (P14).
In contrast, participants who preferred \textit{neck} reported superior calming effects and less distraction, describing it as "more calming" (P4).

\subsubsection{Task Performance and Subjective Response}
\paragraph{\textbf{Task Performance}}

We observed no significant effects of condition on \textbf{task accuracy} ($\chi^{2}(2)=2.886$, $p=0.236$) or \textbf{response time} ($\chi^{2}(2)=5.250$, $p=0.072$). Descriptive statistics were comparable across conditions for accuracy, with \textit{control} ($M=94.76$, $SD=3.29$), \textit{arm} ($M=95.11$, $SD=4.00$), and \textit{neck} ($M=93.49$, $SD=5.16$). 
For response time in milliseconds, the results were \textit{control} ($M=686.92$, $SD=132.18$), \textit{arm} ($M=689.73$, $SD=121.20$), and \textit{neck} ($M=725.09$, $SD=161.58$).

\paragraph{\textbf{Perceived Stress and Affect}}

Self-reported stress, valence, and arousal were descriptively comparable across conditions, suggesting that TES did not produce clear additional subjective benefits over \textit{control}.
Relative to baseline, participants reported increased perceived \textbf{stress}, with \textit{control} ($M=1.46, SD=2.13$), \textit{arm} ($M=1.62, SD=1.95$), and \textit{neck} ($M=1.54, SD=1.72$), decreased \textbf{valence} with \textit{control} ($M=-0.79$, $SD=1.67$), \textit{arm} ($M=-1.62$, $SD=2.55$), and \textit{neck} ($M=-1.08$, $SD=1.56$), and increased \textbf{arousal} with \textit{control} ($M=1.33$, $SD=2.53$), \textit{arm} ($M=1.54$, $SD=1.91$), and \textit{neck} ($M=1.08$, $SD=1.69$).

\subsubsection{Exploratory Physiology Assessment}

For statistical analysis, unless otherwise noted, we employed Friedman’s test to assess overall condition effects, followed by pairwise Wilcoxon signed-rank tests with false discovery rate control using the Benjamini–Hochberg (BH) procedure (FDR = 0.05)~\cite{benjamini_controlling_1995, perneger_whats_1998}.
For each pairwise comparison, we report the effect size $r$ ($=Z/\sqrt{N}$) with its 95\% confidence interval.

\begin{figure}[t]
    \centering
    \includegraphics[width=\linewidth]{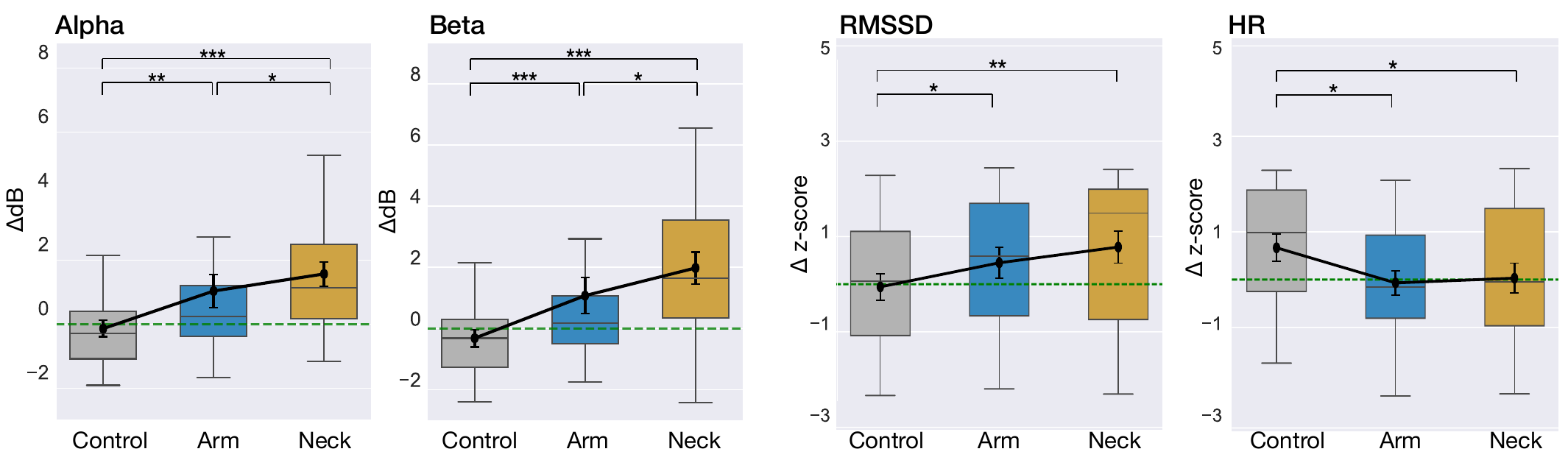}
    \caption{Study~1 exploratory physiological results. 
    Each panel shows three conditions (Control, Arm, Neck). From left to right: baseline-corrected frontal alpha and beta band power changes ($\Delta$ dB), and baseline-corrected RMSSD and HR changes ($\Delta z$-score). 
    Box plots show the distribution of values, and error bars represent mean~$\pm$~SE. Significance markers are shown above comparisons, where * denotes $p<0.05$, ** denotes $p<0.01$, and *** denotes $p<0.001$.}
    \Description{Study~1 physiological results showing EEG and HRV across three conditions (Control, Arm, Neck).}
    \label{fig:study1_result}
\end{figure}

\begin{figure}[t]
    \centering
    \includegraphics[width=\linewidth]{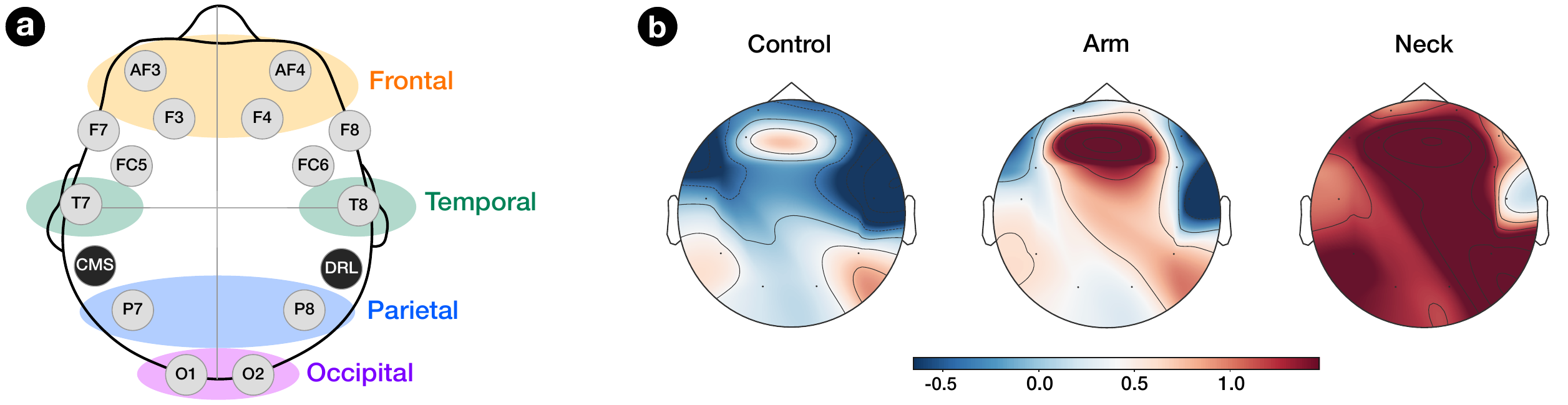}
    \caption{(a) Electrode layout with the EMOTIV EPOC X 14-channel EEG system: frontal (AF3, AF4, F3, F4), temporal (T7, T8), parietal (P7, P8), occipital (O1, O2), and reference (CMS, DRL).
    (b) Topographic distribution of alpha power changes (8-13\,Hz) relative to baseline across conditions. Maps are organized according to the cortical regions, with the frontal area at the top (forehead side), the temporal regions on the left and right (near the ears), the parietal area at the upper back of the head, and the occipital area at the bottom (back of the head). Blue indicates alpha decrease, red indicates alpha increase. The color scale represents $\Delta$\,dB relative to baseline.}
    \Description{Electrode layout and topographic maps of alpha power changes across three Study~1 conditions showing control with reduced alpha, arm with focal frontal increase, and neck with widespread increase.}
    \label{fig:topo}
\end{figure}

\paragraph{\textbf{EEG}}

We observed significant effects of condition across multiple EEG measures as presented in Figure~\ref{fig:study1_result}. 
For \textbf{alpha} power, there was a significant overall effect ($\chi^{2}(2)=15.750$, $p=0.001$). Post-hoc comparisons revealed that both body sites significantly increased alpha power compared to \textit{control}, 
with \textit{arm} ($Z=2.60$, $p=0.009$, $r=0.53$, 95\% CI $[0.16, 0.77]$) and \textit{neck} ($Z=3.83$, $p<0.001$, $r=0.78$, 95\% CI $[0.55, 0.90]$). 
Additionally, \textit{neck} produced significantly higher alpha than \textit{arm} ($Z=2.31$, $p=0.021$, $r=0.47$, 95\% CI $[0.08, 0.73]$).
This pattern is compatible with reduced stress cortical activity, with larger changes at the neck.
For \textbf{beta} power, both sites significantly increased beta power compared to \textit{control}, including \textit{arm} ($Z=3.11$, $p=0.001$, $r=0.63$, 95\% CI $[0.30, 0.82]$) and \textit{neck} ($Z=3.89$, $p<0.001$, $r=0.79$, 95\% CI $[0.57, 0.90]$), with \textit{neck} showing greater increase than \textit{arm} ($Z=2.06$, $p=0.039$, $r=0.42$, 95\% CI $[0.02, 0.70]$).
This pattern was consistent with the alpha results, with \textit{neck} producing stronger neural effects than \textit{arm}, suggesting increased cognitive engagement during the task.

Additionally, we present alpha topographies in Figure~\ref{fig:topo}b to illustrate spatial patterns of stress modulation. 
To visualize these patterns, we computed alpha band power for all 14 channels and averaged baseline-corrected values across participants. The topographic maps show distinct cortical patterns across conditions. The \textit{control} condition showed broadly reduced alpha, especially over frontal regions, consistent with cognitive stress. 
\textit{Arm} was associated with focal increases in frontal alpha power, whereas \textit{Neck} was associated with broader alpha increases across frontal, parietal, and occipital regions. 
These exploratory patterns are consistent with site-dependent differences in physiological responses.

\paragraph{\textbf{HRV}}

For \textbf{RMSSD}, we found a significant overall effect of condition ($\chi^{2}(2)=6.083$, $p=0.048$) as shown in Figure~\ref{fig:study1_result}. Post-hoc comparisons revealed that both TES conditions significantly increased RMSSD compared to \textit{control}, with \textit{arm} ($Z=2.50$, $p=0.019$, $r=0.51$, 95\% CI $[0.13, 0.76]$) and \textit{neck} ($Z=3.02$, $p=0.008$,  $r=0.62$, 95\% CI $[0.29, 0.82]$). No significant difference was found between \textit{arm} and \textit{neck} ($Z=1.38$, $p=0.169$, $r=0.28$, 95\% CI $[-0.14, 0.61]$).
A similar pattern was observed for \textbf{HR}, where a significant overall effect was found ($\chi^{2}(2)=9.083$, $p=0.011$). Both body sites significantly reduced HR compared to \textit{control}, with \textit{arm} ($Z=2.66$, $p=0.024$, $r=0.54$, 95\% CI $[0.17, 0.77]$) and \textit{neck} ($Z=2.21$, $p=0.040$, $r=0.45$, 95\% CI $[0.06, 0.72]$), whereas no significant difference was observed between \textit{arm} and \textit{neck} ($Z=0.38$, $p=0.705$, $r=0.08$, 95\% CI $[-0.33, 0.47]$).
These exploratory findings showed physiological patterns consistent with improved autonomic regulation~\cite{shaffer_overview_2017, rajendra_acharya_heart_2006} relative to the device-free control. However, this could not determine whether the effects were driven by active stimulation, electrode presence, or expectancy.

\section{Study~2: Physiological Assessment of Active Stimulation}

Study~1 established the usability of the multi-node TES system at both arm and neck sites, and exploratory physiological measures suggested stress modulation effects at both placements. 
However, the control condition in Study~1 involved neither stimulation nor electrode attachment, as its focus was the user experience of wearing TES relative to an unencumbered baseline. Although appropriate for that aim, this device-free control meant that the observed physiological differences could not be attributed specifically to active stimulation rather than electrode presence.

We therefore conducted a separate follow-up study with an independent sample, comparing active stimulation with a matched inactive condition at each body site. Unlike Study~1, the electrode pads were attached in all conditions of Study~2, and only the active condition delivered stimulation. This design allowed us to control for electrode presence and expectancy, and to examine whether active stimulation produced site-dependent physiological changes beyond those associated with electrode attachment.

\subsection{Conditions} \label{study2_conditions}
We employed a within-subjects design with four conditions: arm-active (AA), arm-inactive (AI), neck-active (NA), and neck-inactive (NI).
In active conditions, electrotactile stimulation was delivered via electrode pads attached to the arm or neck. In inactive conditions, pads were attached to each site without delivering any stimulation. The calibrated intensities were \textit{arm} ($M=12.32~\mathrm{V}$, $SD = 4.01$) and \textit{neck} ($M = 9.41~\mathrm{V}$, $SD = 1.70$).

\subsection{Participants}
We recruited 24 eligible participants from the local community (9 female and 15 male, $M=21.83~years$, $SD=2.52$), none of whom had participated in Study~1. We set the sample size to match Study~1, in line with participant counts reported in comparable HCI physiological studies~\cite{valente_modulating_2024, paredes_just_2018, balters_calm_2020}.
Eligibility criteria, screening procedures, informed consent, and compensation (\$20 USD) were identical to Study~1.

\subsection{Procedure}

\begin{figure}[t]
    \centering
    \includegraphics[width=\linewidth]{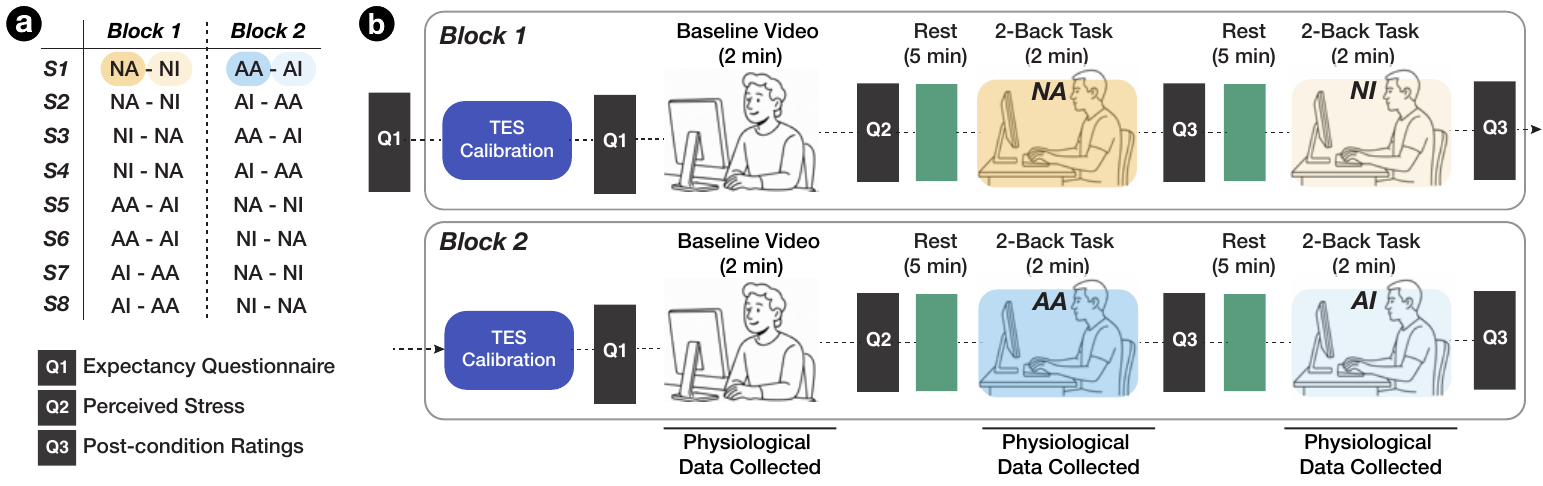}
    \caption{Study~2 procedure: (a) The eight counterbalanced sequences (S1-S8) across two blocks, with questionnaire labels below. 
    (b) Example sequence S1, where Block 1 includes NA--NI and Block 2 includes AA--AI. The session began with an initial expectancy questionnaire (Q1). Conditions were organized into two blocks by site (neck and arm), with the block order fully counterbalanced. Each block began with TES calibration and a post-calibration questionnaire (Q1), followed by a 2-minute baseline video with physiological recording and perceived stress rating (Q2). Within each block, two conditions (active and inactive) were presented in counterbalanced order. Each condition consisted of a 2-minute 2-back task with continuous physiological recording, followed by post-condition ratings (Q3). A 5-minute rest interval separated the two conditions within each block.}
    \Description{Study~2 procedure: (a) The eight counterbalanced sequences (S1-S8) across two blocks, with questionnaire labels below. 
    (b) Example sequence S1, where Block 1 includes NA--NI and Block 2 includes AA--AI. }
    \label{fig:study2_procedure} 
\end{figure}

The task, apparatus, and setup were identical to those in Study~1. 
Because tactile sensation itself was the object of study, complete perceptual blinding was inherently infeasible. Therefore, at the start of the session, participants were informed that the system included modes with varying levels of perceptibility, but were not told which conditions involved no stimulation. We also measured expectancy to assess whether positive bias was present.

The conditions were organized into two blocks: arm (AA and AI) and neck (NA and NI). After attaching electrode pads to the corresponding site, we calibrated TES intensity using the same procedure as Study~1, gradually increasing voltage until a just-perceivable sensation was reported as intensity level 1 on a 1–10 scale.
Participants then completed an expectancy questionnaire regarding the anticipated stimulation experience in stress modulation. 
Each block began with a 2-minute nature video baseline. The two task conditions within each block were separated by a 5-minute rest period.
The procedure is illustrated in Figure~\ref{fig:study2_procedure}, and the order of blocks and conditions within each block was fully counterbalanced across participants.

Upon completing all conditions, participants were debriefed that some conditions included active stimulation while others involved no stimulation, and that this had not been explicitly explained prior to the study to prevent expectation effects. Participants rated the overall study procedure as acceptable ($M=6.63$, $SD=0.58$) on a 7-point Likert scale.

\subsection{Measures}

To assess participants' expectancy of stimulation to help modulate their stress, participants completed a questionnaire before the study on a 7-point Likert scale (1 = not helpful at all, 7 = extremely helpful). 
We also asked participants to rate the same expectancy after calibration at each body site to capture updated expectations following their initial stimulation experience.

Participants rated perceived stress on a 0–10 scale (0 = not stressed at all, 10 = extremely stressed) following each baseline video to provide a pre-condition reference and again following each of the four conditions. After each condition, participants also rated stimulation intensity (1 = not strong at all, 7 = extremely strong) and awareness (1 = not aware at all, 7 = extremely aware).
Perceived stress was computed as the change from the pre-condition baseline to the post-condition rating. Intensity and awareness were included to characterize perceived differences between active and inactive conditions, given that electrotactile stimulation is inherently perceivable. 
Unlike Study~1, which included a full user experience assessment, Study~2 focused on isolating whether active stimulation produced physiological effects beyond those attributable to electrode presence alone.

During the 2-back task and baseline video periods, we measured EEG and HRV as physiological outcomes, consistent with the measures that demonstrated meaningful effects in Study~1. All signals were recorded and processed using the same procedures as Study~1, with each block's baseline video used for baseline correction.

\subsection{Findings}

\subsubsection{\textbf{Physiological Measures}}

For statistical analysis, we conducted two-way repeated measures ANOVAs with site $\times$ stimulation type as within-subject factors, followed by post-hoc paired t-tests with BH-corrected false discovery rate (FDR = 0.05).
We also conducted a sensitivity analysis for the site $\times$ stimulation type interaction using G*Power~\cite{faul_gpower_2007}. With $N=24$, $\alpha=.05$, and power $=0.80$, the design was sensitive to interaction effects of $d_z \ge 0.60$.
For all interaction and post-hoc comparisons, we report effect sizes as $d_z$ with 95\% confidence intervals.

\paragraph{\textbf{EEG}}

\begin{figure}[t]
    \centering
    \includegraphics[width=\linewidth]{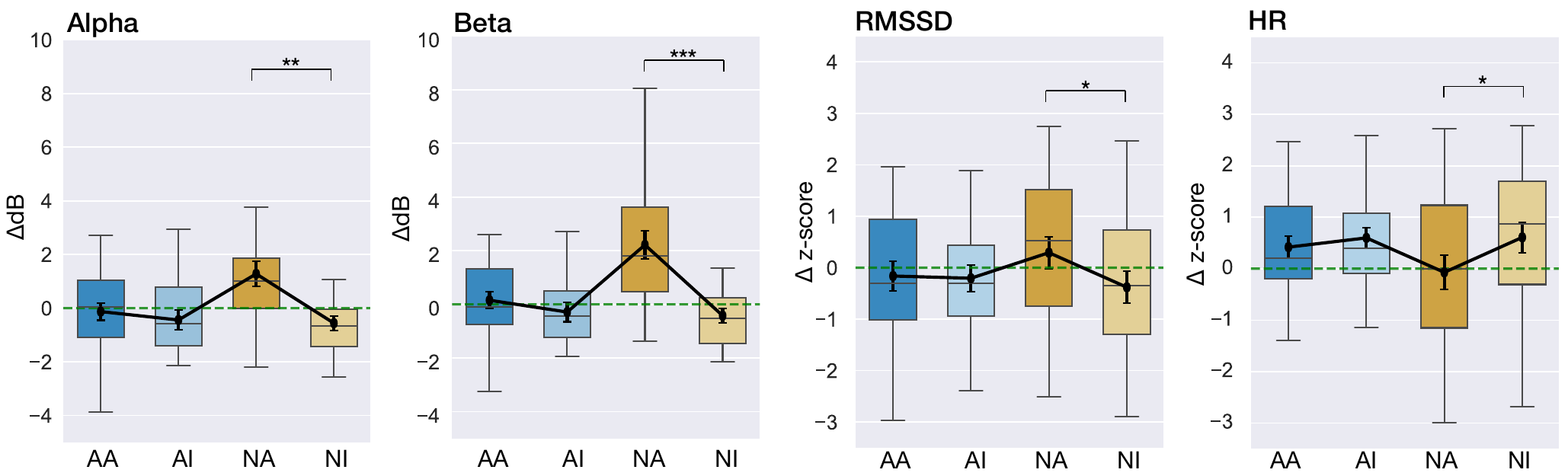}
    \caption{Study~2 physiological results. Each panel shows four conditions (AA, AI, NA, NI). From left to right: baseline-corrected frontal alpha and beta band power changes ($\Delta$ dB), and baseline-corrected RMSSD and HR changes ($\Delta z$-score). 
    Box plots show the distribution of values, and error bars represent mean~$\pm$~SE. Significance markers are shown above comparisons, where * denotes $p<0.05$, ** denotes $p<0.01$, and *** denotes $p<0.001$.}
    \Description{Study~2 physiological results showing EEG and HRV across four conditions (AA, AI, NA, NI).}
    \label{fig:study2_result}
\end{figure}

We found a significant site $\times$ stimulation type interaction for both frontal alpha ($F(1,23)=9.74$, $p=.005$, $d_z=0.64$, 95\% CI $[0.19, 1.07]$) and frontal beta ($F(1,23)=16.74$, $p<.001$, $d_z=0.84$, 95\% CI $[0.36, 1.30]$).
Post-hoc analysis revealed that NA produced significantly greater alpha ($p=.004$, $d_z=0.77$, 95\% CI $[0.31, 1.22]$) and beta power ($p<.001$, $d_z=0.96$, 95\% CI $[0.47, 1.44]$) than NI, whereas AA and AI did not differ significantly for either alpha ($p=.313$, $d_z=0.25$, 95\% CI $[-0.16, 0.65]$) or beta measure ($p=.141$, $d_z=0.34$, 95\% CI $[-0.07, 0.75]$), as shown in Figure~\ref{fig:study2_result}.

\begin{figure}[t]
    \centering
    \includegraphics[width=0.75\linewidth]{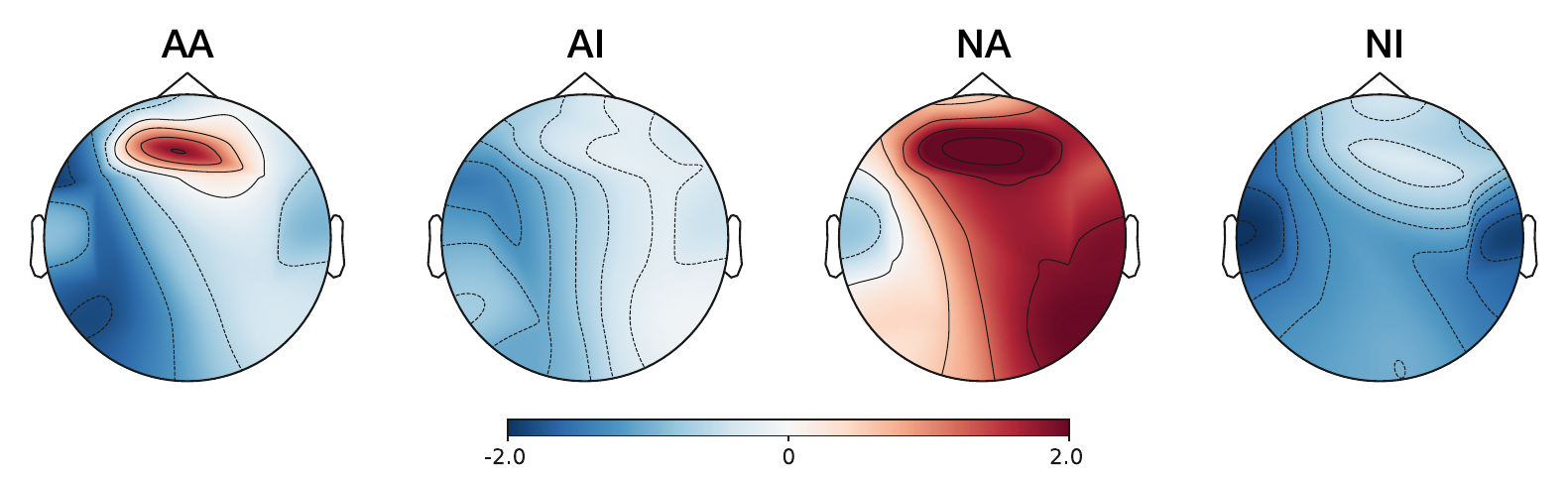}
    \caption{Topographic distribution of alpha power changes (8-13\,Hz) relative to baseline across Study~2 conditions. Blue indicates alpha decrease, red indicates alpha increase. Visual patterns reflect group means, and NA shows widespread alpha increases. The color scale represents $\Delta$\,dB relative to baseline.}
    \Description{Topographic maps of alpha power changes across four Study~2 conditions. Arm active, arm inactive, and neck inactive show blue (alpha decrease), while neck active shows widespread red (alpha increase).}
    \label{fig:topo2}
\end{figure}

We also computed alpha topographic maps as shown in Figure~\ref{fig:topo2} for all 14 channels using the same approach as Study~1. NA showed widespread alpha increases, while AI and NI showed comparable patterns of task-related alpha suppression, consistent with the statistical results. AA showed a focal increase at a single frontal channel, but this did not reach significance.

\paragraph{\textbf{HRV}}
For RMSSD, we observed a significant site $\times$ stimulation type interaction ($F(1,23)=4.67$, $p=.041$, $d_z=0.44$, 95\% CI $[0.02, 0.86]$).
Post-hoc comparisons revealed that NA significantly increased RMSSD compared to NI ($p=.021$, $d_z=0.63$, 95\% CI $[0.19, 1.06]$), whereas AA and AI did not differ ($p=.987$, $d_z=0.06$, 95\% CI $[-0.34, 0.46]$).
For HR, we observed a significant site $\times$ stimulation type interaction ($F(1,23)=6.19$, $p=.020$, $d_z=-0.51$, 95\% CI $[-0.93, -0.08]$). 
Post-hoc comparisons showed that NA significantly reduced HR compared to NI ($p=.016$, $d_z=-0.66$, 95\% CI $[-1.09, -0.21]$), whereas AA and AI did not differ ($p=.475$, $d_z=-0.15$, 95\% CI $[-0.55, 0.26]$), consistent with the RMSSD pattern (see Figure~\ref{fig:study2_result}).

\subsubsection{\textbf{Task Performance}}
Given the high mean 2-back task accuracy (M $>$ 90\%), we used nonparametric Friedman tests to analyze task performance, consistent with Study~1. 
No significant condition effects were found for accuracy ($\chi^{2}(3)=5.34$, $p=.149$) or response time ($\chi^{2}(3)=1.35$, $p=.717$).
Descriptive statistics were comparable across conditions for accuracy, with AA ($M=93.38$, $SD=5.33$), AI ($M=92.16$, $SD=5.71$), NA ($M=90.47$, $SD=6.58$), and NI ($M=93.70$, $SD=5.17$). For response time in milliseconds, the results were AA ($M=696.98$, $SD=152.03$), AI ($M=699.83$, $SD=183.20$), NA ($M=719.81$, $SD=136.96$), and NI ($M=707.59$, $SD=149.80$).

\subsubsection{\textbf{Subjective Measures}}

We report subjective measures descriptively to contextualize expectancy, perceived stimulation, and perceived stress across active and inactive conditions.

\paragraph{\textbf{Expectancy}}
Prior to the experiment, participants reported neutral expectations regarding stimulation effectiveness ($M=4.33$, $SD=1.27$). 
Following calibration, post-calibration expectancy remained near or below neutral for both sites, with arm ($M=3.29$, $SD=1.16$) and neck ($M=3.71$, $SD=1.30$), suggesting that participants did not enter the task with strong positive expectations about stimulation effectiveness.

\paragraph{\textbf{Perceived Experience}}
Participants reported detecting stimulation more readily in active conditions at both sites.
For the arm, AA was rated higher in intensity ($M=2.39$, $SD=1.59$) and awareness ($M=3.91$, $SD=2.06$) compared to AI, where participants still reported non-negligible intensity ($M=1.74$, $SD=1.25$) and awareness ($M=2.26$, $SD=1.63$) despite receiving no stimulation. 
A similar pattern was observed at the neck, where NA was rated higher in intensity ($M=3.17$, $SD=1.96$) and awareness ($M=4.04$, $SD=2.14$), while NI also remained above the scale minimum in intensity ($M=2.33$, $SD=1.61$) and awareness ($M=3.08$, $SD=2.38$). These non-negligible ratings in inactive conditions suggest that pad presence alone produced some perceived sensation.
%
Additionally, baseline-corrected perceived stress was comparable across active and inactive conditions at both sites. For the arm, AA ($M = 0.92$, $SD = 1.35$) and AI ($M = 0.96$, $SD = 1.49$) showed descriptively similar values of perceived stress. For the neck, NA ($M = 1.25$, $SD = 1.87$) and NI ($M = 0.67$, $SD = 1.97$) also did not differ meaningfully.

\section{DISCUSSION}

\subsection{Neck TES Was Acceptable Despite Site Sensitivity} \label{discussion1}

Addressing \textbf{RQ1}, our results show that multi-node TES was acceptable on the neck despite the site's tactile sensitivity and social visibility.
A key concern for deploying TES on the neck was that electrotactile stimulation is often associated with unpleasant sensations~\cite{delitto_study_1992, faltaous_understanding_2024}. Surprisingly, our results suggest otherwise. 
Neck stimulation at a sensitive and socially visible site was perceived as acceptable during cognitively demanding tasks. Despite its location, the electrotactile experience on the neck was rated as comfortable and pleasant by most participants, with future use intentions comparable to the arm. 
Given that the neck is also the site where physiological stress modulation was observed, these findings directly support that the neck should remain in the design space for wearable TES, rather than being dismissed due to presumed discomfort or social sensitivity.

The two sites evoked qualitatively distinct experiences. 
Arm stimulation was described in familiar, concrete terms such as tapping or buzzing, whereas neck stimulation elicited more varied and metaphorical descriptions, including raindrops, butterflies, and gentle throbbing reminiscent of bodily rhythms such as heartbeats. This resonance with intrinsic bodily rhythms may help explain why some participants perceived the neck as simultaneously more intrusive and more calming. The arm offered practical advantages familiar from everyday wearables, including easier concealment and lower perceived intrusiveness, which remain relevant for form factor decisions even though arm stimulation did not modulate physiology.

Some participants described TES on the arm as resembling a comforting presence, such as someone tapping, holding, or reassuring them. Although electrotactile stimulation primarily activates A$\beta$ afferents rather than CT afferents~\cite{ackerley_c-tactile_2022}, the slow spatiotemporal movement pattern may have been cognitively interpreted as social touch~\cite{jiang_designing_2024, nisky_neurocognitive_2024}, even though electrotactile stimulation does not directly engage CT afferents.

\subsection{Physiological Stress Modulation Depended on Placement}

Our Study~1 showed promising EEG and HRV changes at both arm and neck sites relative to a device-free control. Because Study~1 focused on comparing subjective user experience against a device-free baseline, the control did not include electrode attachment. Study~2, therefore, separated stimulation effects from electrode presence by comparing active and inactive stimulation under identical attachment at each site.
We found the neck and arm sites dissociated. 
Relative to NI, NA produced robust cortical changes and more moderate but consistent autonomic changes, whereas AA did not differ from AI.

Neck stimulation produced robust increases in both frontal alpha and beta power. These concurrent increases are consistent with a state of calm alertness, where relaxation coexists with maintained cognitive engagement~\cite{sugimoto_electroencephalographic_2024}. A similar co-elevation of alpha and beta band power has been observed in meditation states~\cite{attar_eeg-based_2025, duda_mindfulness_2024}. Furthermore, these cortical changes were accompanied by increased RMSSD and reduced HR in our studies, a pattern consistent with increased parasympathetic regulation~\cite{shaffer_overview_2017, rajendra_acharya_heart_2006}. 
Together, the coordinated cortical and autonomic changes suggest physiological stress modulation at the neck placement and are consistent with findings from prior vagal TES studies. However, because we did not directly measure vagal engagement, these changes cannot be attributed specifically to vagal activation. Participants' experiences of the stimulation, discussed in \S~\ref{discussion1}, may also have contributed to the observed responses, suggesting that physiological and experiential processes may have operated concurrently. 
While the tactile nature of the stimulation makes these contributions difficult to fully disentangle, our findings support the feasibility of applying the spatiotemporal configuration evaluated here at the neck while maintaining user acceptability.

The arm findings clarify how the Study~1 exploratory effects should be interpreted.
Because AA did not differ detectably from AI under identical electrode attachment, the Study~1 arm effects cannot be attributed specifically to active electrotactile stimulation. They may instead have included non-specific factors associated with the device-free comparison, such as pad presence, bodily attention, or expectancy.
Non-negligible intensity and awareness ratings in inactive conditions support this interpretation, showing that electrode attachment can generate perceived bodily sensations. 
Addressing \textbf{RQ2}, our original rationale was that electrotactile feedback might modulate stress similarly to vibrotactile feedback at a common wearable site. However, measurable physiological differences between the active and inactive conditions emerged only at the neck, even though participants received the same TES stimulation pattern at both body sites and rated it as similarly acceptable and pleasant. 
Thus, our findings did not support the proposed analogy between electrotactile and vibrotactile feedback for stress modulation at the arm, under the configuration tested here.

\subsection{Physiological Changes Did Not Translate Into Subjective Awareness}

Our system was designed for tactile comfort and stress modulation rather than subjective affect modulation. Although the neck TES produced physiological stress modulation, we found no corresponding changes in perceived stress. Such dissociations between physiological and subjective stress markers are commonly reported~\cite{shah_unveiling_2024, paredes_just_2018}, and several candidate explanations are consistent with our data.

First, cognitive load during the 2-back task could have suppressed interoception, the processing of internal bodily signals, preventing conscious perception of physiological changes~\cite{price_interoceptive_2018}.
Second, physiological signals were recorded continuously, whereas perceived stress ratings were collected retrospectively. This delay, combined with the cognitive load, likely reduced the sensitivity of self-reports to capture subtle changes.
Third, participants described TES as low-salience, suggesting stress modulation may have operated near or below the threshold of conscious awareness, where physiological changes do not necessarily translate into reportable subjective shifts~\cite{winkielman_unconscious_2004}. 
Fourth, a more speculative possibility involves limited CT afferent engagement. Prior work suggests that electrical stimulation at comfortable cutaneous levels is unlikely to directly activate CT afferents~\cite{ackerley_c-tactile_2022}. 
CT afferents play a role in translating tactile input into conscious feelings of comfort or calm. As our TES was delivered through the skin, limited CT engagement may help explain why physiological changes did not translate into corresponding subjective reports.

From an HCI perspective, this dissociation has practical implications. Without subjective cues of effectiveness, users may struggle with long-term adherence~\cite{murnane_mobile_2015}. Feedback mechanisms that make physiological improvements visible, such as HRV trends or progress indicators, could help bridge this gap while preserving the low-salience operation that makes TES unobtrusive in daily use~\cite{stetler_adherence_2014}.

\subsection{TES Did Not Interfere with Cognitive Performance}

Task performance on the brief 2-minute task was maintained across all conditions in both studies, regardless of site or stimulation state. 
For the neck, where physiological stress modulation was observed, maintained performance alongside reduced physiological load suggests that TES can reduce stress without imposing a performance cost~\cite{mahdavi_unraveling_2024, mandrick_neural_2016}. 
For conditions without physiological effects, stable performance was also observed. 
This indicates that the multi-node TES did not disrupt ongoing cognitive tasks, consistent with prior work on low-salience, ambient haptic cues that can be integrated without compromising focus or accuracy~\cite{whitmore_improving_2024}.

Qualitative reports during the usability study showed varied attentional responses, as participants described both divided attention and enhanced focus. Neither pattern translated into measurable performance changes, suggesting that mild awareness of stimulation can coexist with stable task performance, rather than functioning as a distractor. 
This outcome is especially valuable in applied settings as TES can be deployed in contexts requiring sustained mental engagement without imposing a performance penalty, a prerequisite for any stress modulation intervention intended for workplace or study settings~\cite{stansfeld_psychosocial_2006, howe_design_2022}.

\section{IMPLICATIONS}

We derive three design implications for TES stress modulation from our study findings.

\paragraph{\textbf{Separate physiological efficacy from tactile comfort}}

For TES systems intended to modulate physiological stress, tactile comfort may not be sufficient. In our studies, arm and neck stimulation were both rated as comfortable and acceptable, yet physiological effects beyond inactive stimulation emerged only at the neck. 
This suggests that TES design should evaluate placement along two distinct dimensions: whether the site is physiologically relevant to the intended mechanism, and whether the resulting sensation is acceptable for stress modulation use. 
Electrotactile feedback can still support comfortable and interpretable tactile experiences, but designers should not assume that tactile acceptability at a given site implies physiological stress modulation at that site.

\paragraph{\textbf{Design for acceptability at sensitive body sites}}

Participants rated multi-node TES on the neck as comparably comfortable and acceptable to the arm, suggesting that the neck should not be dismissed as an impractical deployment site solely because it is sensitive or socially visible.
Despite initial concerns about sensitivity, most participants found neck stimulation reasonably comfortable and expressed willingness for future use comparable to the arm. These findings open design opportunities for neck-worn TES systems, such as skin-tone matched patches, collar-integrated designs, or jewelry-like wearables that reduce social visibility while preserving access to physiologically relevant sites~\cite{euler_review_2022, greig_electrical_2024}.

\paragraph{\textbf{Consider body-site differences in sensory interpretation}}
Our findings suggest that body site shapes not only comfort but also how stimulation is interpreted.
Participants described the same stimulation differently depending on the body site. Neck stimulation was described through metaphors drawn from natural phenomena and bodily rhythms, such as raindrops, butterflies, and heartbeat-like throbbing, whereas arm stimulation was often interpreted through familiar tactile experiences such as tapping, pulsing, or being held. 
These responses suggest that the same spatiotemporal stimulation pattern may be interpreted differently across body sites. Designers should therefore consider how a given pattern is perceived at different placements.

\section{LIMITATIONS \& FUTURE WORK}

We employed 2-minute stimulation blocks as a conservative safety measure given limited guidance for extended vagal stimulation~\cite{howland_vagus_2014, yap_critical_2020}. Future work should employ longer measurement periods and consider additional autonomic markers such as pupil dilation~\cite{lowenstein_role_1950} to strengthen our findings. 
Furthermore, a sensitivity analysis indicated that Study~2 could detect moderate-to-large within-participant interaction effects. Because the autonomic effects were smaller and estimated with wider confidence intervals, future studies with larger samples could provide more precise estimates of these effects.

Our design targeted electrotactile sensations in a comfortable range through multi-node distribution and individually calibrated just-perceptible intensity. A direct comparison across a range of intensity levels (e.g., subliminal, perceptible, salient)~\cite{elgendi_subliminal_2018} was beyond the scope of this work. Future studies could systematically examine how perceptibility shapes both physiological and subjective outcomes, clarifying whether conscious perception of stimulation is necessary for, or incidental to, the stress modulation effects observed at the neck.
Relatedly, our voltage-controlled stimulation, though practical for wearables, may be less robust to impedance changes than current-controlled stimulation. In our setting, per-participant, per-site calibration mitigated baseline impedance differences, but a direct comparison of the two approaches remains valuable future work~\cite{vance_skin_2015}.

Although we used a fixed spatiotemporal pattern, our multi-node TES system enables broader design possibilities. As discussed, some participants described arm TES as resembling a comforting presence, suggesting that pattern design may shape the cognitive interpretation of electrotactile feedback even without physiological stress modulation. Future work could explore whether spatiotemporal TES patterns can be designed to function as social affective cues, for example, by approximating familiar comforting touch experiences, and how parameters such as speed, direction, and rhythm shape perceived comfort and social presence.
In addition, given that vibrotactile feedback has been shown to support stress modulation~\cite{miri_evaluating_2020}, electrotactile feedback may still do so under other parameters, intensities, or body placements, which remains an important direction for future work.

Finally, our findings are limited to cognitive stress contexts, and TES effectiveness may vary across social, environmental, or physical stressors and intensities. Future work should investigate TES efficacy in more naturalistic settings and compare it against other haptic modalities, such as vibrotactile, pressure-based, and thermal feedback, to assess its relative advantages in both efficacy and deployment viability for stress management.

\section{CONCLUSION}

This study investigated TES for stress modulation at two body sites: the \textit{neck}, a placement used in prior vagal TES work, and the \textit{arm}, which primarily delivers electrotactile sensation.
Through a usability study followed by a controlled physiological assessment, we found that neck TES produced cortical and autonomic changes consistent with stress modulation, whereas arm TES did not differ from inactive conditions on any physiological measure.
These results highlight the importance of placement in TES stress modulation, as physiological changes emerged at the neck, whereas arm stimulation did not differ from inactive electrode attachment under the tested configuration.
Importantly, participants rated neck stimulation as comparably comfortable and acceptable to the arm with our multi-node TES, suggesting that the neck should not be dismissed as an impractical deployment site despite its sensitivity. 
Although physiological stress modulation was observed, participants did not report corresponding changes in perceived stress, suggesting that physiological effects of low-salience TES may not be consciously perceived. Task performance remained stable across all conditions in both studies, indicating that the TES did not interfere with the brief cognitive task used in our studies.
These findings provide a foundation for deploying TES as a stress modulation system, with design considerations for placement at physiologically relevant sites, acceptability at sensitive body sites, and sensory interpretation of electrotactile patterns.

\begin{acks}
We thank Dr. Caitlin Ridgewell for her guidance on the neurophysiological mechanisms of electrical stimulation, Dr. Myungin Lee for access to the EEG device, and Dr. Jun Nishida for advice on circuit design. We also thank Shalom Akpakla for assistance with study sessions, all study participants for their time and participation, and the reviewers for their constructive feedback. Claude and ChatGPT were used to assist with manuscript editing, including proofreading, improving clarity, and increasing concision.
\end{acks}

\bibliographystyle{ACM-Reference-Format}
\bibliography{references}

\end{document}